\documentstyle[twoside,aps,prl,multicol,epsf]{revtex}
\newcommand\lsim{\lower 3pt\hbox{$\buildrel < \over\sim$}}
\newcommand\gsim{\lower 3pt\hbox{$\buildrel > \over\sim$}}
\newcommand\lessgtr{\lower 3 pt\hbox{$\buildrel < \over > $}}
\begin{document}
\title{Decay laws for three-dimensional magnetohydrodynamic turbulence}
\author{Dieter Biskamp and Wolf--Christian M\"uller}
\address{Max-Planck-Institut f\"ur Plasmaphysik,\\
85748 Garching, Germany }
\draft
\maketitle
\begin{abstract}
Decay laws for three-dimensional incompressible magnetohydrodynamic turbulence
are obtained from high-resolution numerical simulations using up to $512^3$ modes.
For the typical case of finite magnetic helicity $H$ the energy decay is found
to be governed by the conservation of $H$ and the decay of the energy ratio 
$\Gamma=E^V/E^M$. One finds the relation $(E^{5/2}/\epsilon H)\Gamma^{1/2}/
(1+\Gamma)^{3/2}=const$,  $\epsilon =-dE/dt$. Use of the observation that
$\Gamma(t)\propto E(t)$ results in the asymptotic law $E\sim t^{-0.5}$
in good agreement with the numerical behavior. For the special case $H=0$ the energy
decreases more rapidly $E\sim t^{-1}$, where the transition to the finite-$H$ behavior
occurs at relatively small values.
\end{abstract}


\begin{multicols}{2}

Many plasmas, especially in astrophysics, 
are characterized by turbulent magnetic fields,
the best-known and most readily observable example being the solar wind.
The convenient framework to describe such turbulence is magnetohydrodynamics (MHD).
Here one ignores the actual complicated dissipation processes, 
which occur on the smallest scales and would usually
require a kinetic treatment, assuming that the main turbulent scales are
essentially independent thereof. Instead dissipation is modeled by simple 
diffusion terms. If, moreover, interest is focussed on the intrinsic turbulence dynamics,
one can also ignore the largest scales in the system, which depend on
the specific way of turbulence generation, restricting consideration to a
small open homogeneous domain of the globally inhomogeneous turbulence. 

Homogeneous MHD turbulence has become a paradigm in fundamental
turbulence research, which has been receiving considerable attention. 
It is well known that 2D and 3D MHD turbulence have many features in common concerning,
in particular, the cascade properties. In both cases there are three quadratic
ideal invariants: the energy $E=\frac{1}{2}\int(v^2+B^2)dV$, 
the cross helicity $K=\int {\bf v}\cdot {\bf B} dV$, and a purely magnetic quantity,
the magnetic helicity $H=\int {\bf A}\cdot{\bf B} dV$ in 3D and the mean-square
magnetic potential $H^{\psi}=\int \psi^2 dV$ in 2D, which both exhibit an inverse
cascade. Many theoretical predictions do not distinguish between 2D and 3D, 
concerning, e.g.,  the tendency toward velocity and magnetic field alignment or the 
spectral properties. Thus it is not surprising, that numerical studies of MHD turbulence
have been mostly concentrated on two-dimensional simulations,
where high Reynolds numbers can be
reached much more readily, see e.g., \cite{1}--\cite{6}.   
While 2D simulations are now being performed with up to 
$N^2 =4096^2$ modes (or, more accurately,
collocation points)
\cite{6}, studies of 3D MHD homogeneous turbulence have to date been restricted
to relatively low Reynolds numbers using typically $N^3=64^3$ modes, 
e.g., \cite{7}, \cite{8}, precluding an inertial range
scaling behavior. Also in Ref.\ \cite{9}, where a somewhat higher Reynolds number could be 
reached by using $180^3$ 
modes, attention was focussed primarily on the process of
turbulence generation from smooth initial conditions and the properties of the prominent
spatial structures, current and vorticity sheets. 

In this Letter we present results of a numerical study of freely decaying
3D MHD turbulence with spatial resolution up to $512^3$ modes. We discuss the
decay laws of the integral quantities, in particular the energy $E$
and the ratio of kinetic and magnetic energies $\Gamma = E^V/E^M$, and
their dependence on the quasi-constant value of $H$. The energy decay is found
to follow a simple law, which
is determined by $\Gamma(t)$ and $H$. While most previous studies have been
restricted to the case of negligible magnetic helicity $H\simeq 0$, we focus
attention on the properties of the turbulence for finite $H$, which is more typical
for naturally existing MHD turbulence occuring mostly in rotating systems.
We find that for finite $H$ the energy decays significantly more slowly than for
$H\simeq 0$. This behavior is primarily caused by the rapid decrease of the
energy ratio $\Gamma$, which has the same decay time as the energy.

The 3D incompressible MHD equations, written in the usual units,  
\begin{equation}
\partial_t{\bf B} -\nabla\times({\bf v}\times{\bf B})=\eta_{\nu}(-1)^{\nu-1}
\nabla^{2\nu}{\bf B},
\label{1}
\end{equation}
\begin{equation}
\partial_t{\bf w} -\nabla\times({\bf v}\times{\bf w})-\nabla \times({\bf j}\times
{\bf B})=\mu_{\nu}(-1)^{\nu-1}\nabla^{2\nu}{\bf w},
\label{2}
\end{equation}
\[
{\bf w} = \nabla \times {\bf v}, \quad {\bf j}=\nabla\times{\bf B},
\]
are solved 
in a cubic box of size $2\pi$ with periodic boundary
conditions. The numerical method is
a pseudo-spectral scheme with spherical mode truncation as conveniently
used in 3D turbulence simulations (instead of full dealiasing by the 2/3 rule
chosen in most 2D simulations). 
Initial conditions are
\begin{equation}
{\bf B}_{\bf k} = a\, {\rm e}^{-k^2/k^2_0-i\alpha_{\bf k}},
\quad {\bf v}_{\bf k} = b\, {\rm e}^{-k^2/k^2_0-i\beta_{\bf k}},
\label{2a}
\end{equation}
which are characterized by random phases $\alpha_{\bf k}$, $\beta_{\bf k}$ and
satisfy the conditions ${\bf k}\cdot{\bf B}_{\bf k}={\bf k}\cdot{\bf v}_{\bf k}=0$
as well as
$E=1$ and $\Gamma=1$. Further restrictions on ${\bf B}_{\bf k}$ and
${\bf v}_{\bf k}$ arise by requiring specific values of $H$ and $K$, respectively.
The wavenumber $k_0$, the location of the maximum of the initial energy spectrum,
is chosen as $k_0=4$, which allows the inverse cascade of $H_{\bf k}$ to develop
freely during the simulation time of 10-20 eddy turnover times.
This implies 
a certain loss of inertial range, i.e., a reduction in Reynolds number, but
the sacrifice is unavoidable in the presence of inverse cascade dynamics.
Choosing $k_0\sim 1$ would lead to magnetic condensation in the lowest-$k$ state,
which would affect the entire turbulence dynamics.
We have used both normal diffusion $\nu=1$ and hyperdiffusion $\nu=2$.
Apart from the fact that inertial ranges are wider and $H$ is much better conserved
for $\nu=2$ than for $\nu=1$, no essential differences are found between the two cases.
The generalized magnetic Prandtl number
$\eta_{\nu}/\mu_{\nu}$ has been set equal to unity.
Table \ref{t1} lists the most important parameters of the simulation runs.
 
The energy decay law is a characteristic property of a turbulent system.
In hydrodynamic turbulence  the decay rate depends on the energy spectrum at
small $k$. Assuming time invariance of the Loitsianskii integral 
${\cal L}=\int^{\infty}_0 dl\, l^4\langle v_l(x+l)v_l(x)\rangle$ 
the energy has been  predicted to
follow the similarity law $E\sim t^{-10/7}$ \cite{10}.
The invariance of ${\cal L}$ has, however, been  questioned, see e.g., \cite{11}. Both
closure theory \cite{12} and low-Reynolds number simulations \cite{8} yield  
a significantly slower decrease, $E\sim t^{-1}$. Experimental measurements of the energy
decay law $t^{-n}$ are rather difficult and do not give a uniform picture, $n$
ranging between 1.3 \cite{21} and 2 \cite{22}. 

The invariance of the Loitsianskii integral has recently also been postulated for
MHD turbulence \cite{4}, where ${\cal L}_{\rm MHD}$ is defined in analogy to
${\cal L}$ in terms of the longitudinal correlation function 
$\langle z^{\pm}_l(x+l)z^{\pm}_l(x)\rangle$ of the Elsaesser fields
${\bf z}^{\pm}={\bf v}\pm {\bf B}$. Since $z^2\sim E$, this assumption gives
${\cal L}_{\rm MHD} \sim L^5 E =const$, where $L$ is the macroscopic scale length
of the turbulence. In addition
the expression for the energy transfer $dE/dt =-\epsilon \sim -z^4/LB_0$ was used,
which formally accounts for the Alfv\'en effect \cite{14},\cite{15}. 
These relations give  $(dE/dt)B_0/E^{11/5}=const$ and hence $E\sim t^{-5/6}$,
treating $B_0$ as constant.
One may, however, argue that the Alfv\'en effect is only
important on small scales $l\ll L$, while on the scale $L$ of the energy-containing
eddies $B_0$ is not constant but $B_0\sim E^{1/2}$
(except for the case that $B_0$ is an  external field, which would,
however, make the turbulence strongly
anisotropic), hence $\epsilon \sim E^{3/2}/L$,
which would give  the same result $n=10/7$ as predicted for hydrodynamic turbulence. 
Low-resolution numerical simulations \cite{8}
indicate $n\simeq 1$, which is also found in recent simulations of compressible
MHD turbulence \cite{23}.

For finite magnetic helicity $H$ provides a constant during energy
decay, which for high Reynolds number is more
robust than the  questionable invariance of the Loitsianskii
integral. It is true that in contrast to the 2D case, where $E^M$ and $H^{\psi}$
are tightly coupled, such that $E^M\neq 0$ implies $H^{\psi}\neq 0$, in 3D
a state with $H=0$ and finite magnetic energy is possible. But this is only
a special and not typical case, since in nature magnetic turbulence usually
occurs in rotating systems, which give rise to finite magnetic helicity.
   
\begin{figure}
\epsfxsize=9truecm
\epsfbox{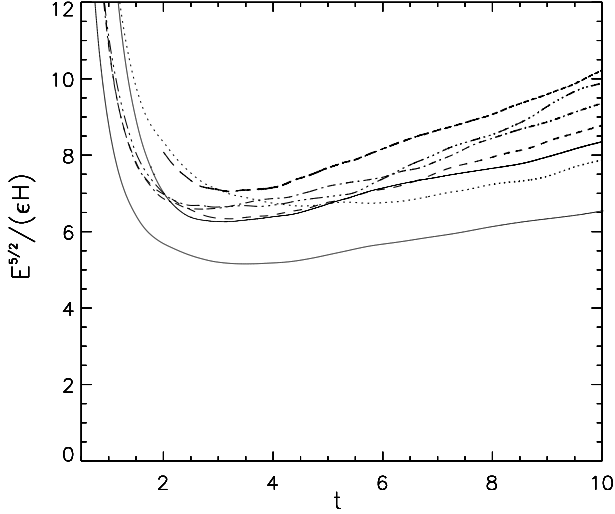}
\caption{Energy decay law, displayed in the differential form 
 $E^{5/2}/(\epsilon H)$ for the runs 1,2,4,5,6,7,8 
in Table 1. The increase in time indicates an energy decrease
slower than $t^{-2/3}$, typically $t^{-0.5}$.} 
\label{f1a}
\end{figure}

If the process of turbulence decay is self-similar, which also implies 
that the energy ratio $\Gamma$ remains constant, 
the energy decay law follows from a simple argument \cite{13}. With
the scale length $L=E^{3/2}/\epsilon$, the dominant scale of the energy-containing
eddies,  
we have 
\begin{equation}
H \simeq E^ML \sim EL,
\label{2b}
\end{equation}
since owing to the assumed self-similarity $E^M \sim E^V \sim E$. Inserting $L$ gives
\begin{equation}
-\frac{dE}{dt} = \epsilon \sim \frac{E^{5/2}}{H},
\label{3}
\end{equation}
which has the similarity solution $E\sim t^{-2/3}$. 
In Fig. \ref{f1a} the ratio $E^{5/2}/(\epsilon H)$ is plotted for 
the runs from Table \ref{t1} with $H\neq 0$ and small initial correlation 
$\rho_0$.
The figure shows that this quantity is
not constant, but increases in time. Moreover, there is a significant
scatter of the different curves. Integration yields
a slower asymptotic energy decay than predicted $n\simeq 0.5-0.55$.     
(The log-log representation
of $E(t)$, often given in the literature to make a power law behavior visible,
is misleading, since the major part of such a curve refers to the transition period
of turbulence generation. The solution $(t-t_*)^{-n}$ approaches the power law
$t^{-n}$ only asymptotically for $t\gg t_*$, where $t_*$ is
not accurately known.
We therefore prefer to plot the decay law in the primary differential form.)

\begin{figure}
\epsfxsize=9truecm
\epsfbox{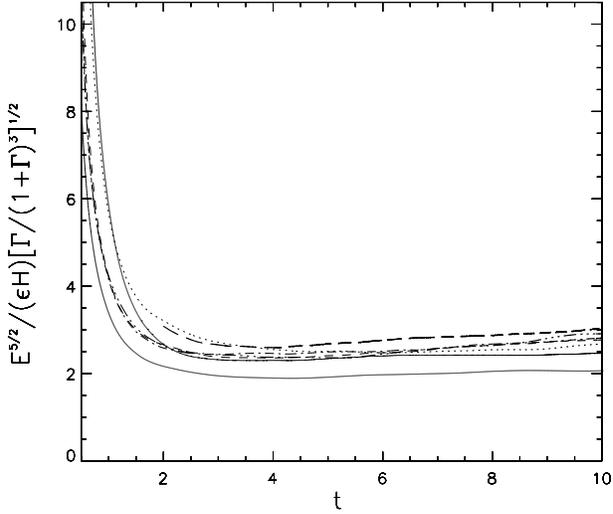}
\caption{Energy decay law in differential form 
$(E^{5/2}/\epsilon H) \Gamma^{1/2}/
(1+\Gamma)^{3/2}$ for the same runs as in Fig.\ \ref{f1a}. The lowest curve,
which falls somewhat outside the main curve bundle, corresponds
to the run with the smallest Reynolds number (run 1), where conservation of $H$ is least good. }
\label{f1b}
\end{figure}

We can attribute this discrepancy to the fact that the turbulence does not decay  
in a fully self-similar way. Indeed  the energy ratio $\Gamma$ is 
found to decrease rapidly,
in contrast to the 2D case,
where $\Gamma$ decays much more slowly, typically logarithmically 
\cite{1}, \cite{3}. (The ratio of viscous and resistive
dissipation $\epsilon^{\mu}/\epsilon^{\eta}$, however, 
remains constant just as in the 2D case \cite{3}, 
which simply reflects the basic property, that
dissipation takes place in current sheets and that these are also vorticity sheets, i.e.,
the location of viscous dissipation.)
Let us incorporate the dynamic change 
of $\Gamma$ in the theory of the energy decay. Assuming that the most important 
nonlinearities arise from the ${\bf v}\cdot\nabla$ contributions in the MHD
equations, Eq.\ (\ref{3})
is replaced by
\begin{equation}
\epsilon \sim (E^V)^{1/2}\frac{E}{L} = 
\frac{\Gamma^{1/2}}{(1+\Gamma)^{3/2}}
\frac{E^{5/2}}{H},
\label{4}
\end{equation}
using the relation (\ref{2b}). Figure \ref{f1b} shows that $(E^{5/2}/\epsilon H)\Gamma^{1/2}/
(1+\Gamma)^{3/2}$ is indeed nearly constant for $t>2$, when turbulence is fully developed,
and the scatter in Fig.\ \ref{f1a} is strongly
reduced. Hence relation (\ref{4}) is generally valid  for finite magnetic
helicity. It is also independent of the magnitude
of the dissipation coefficients and character
of the dissipation ($\nu=1$ or 2), as long as $H$ is well conserved.

\begin{figure}
\epsfxsize=9truecm
\epsfbox{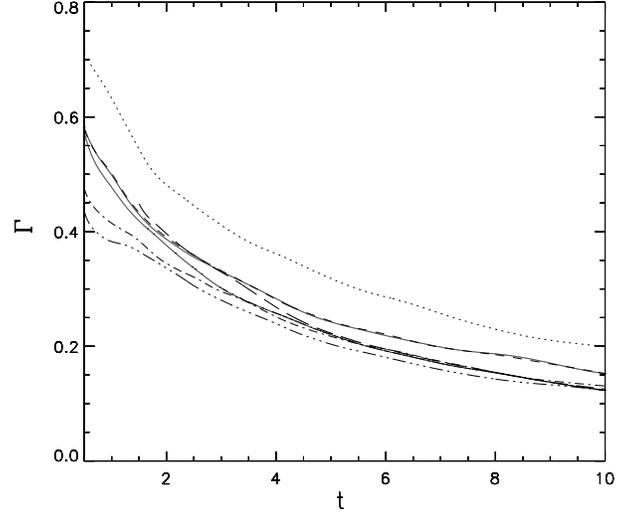}
\caption{Energy ratio $\Gamma(t)$ for the same runs
as in Fig.\ \ref{f1a}.}
\label{f2}
\end{figure}

\begin{figure}
\epsfxsize=9truecm
\epsfbox{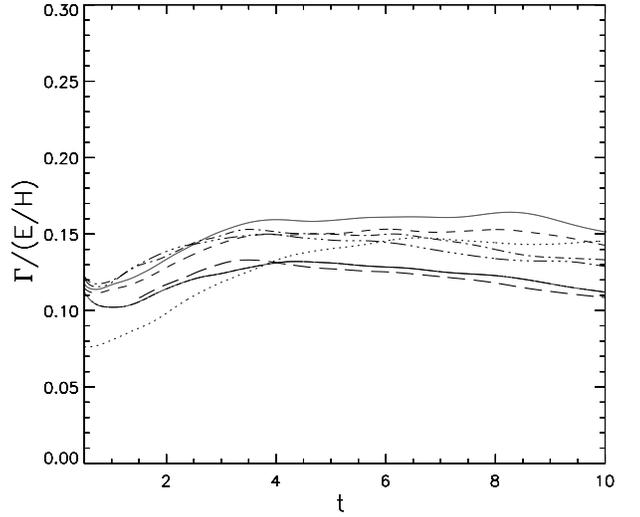}
\caption{$\Gamma/(E/H)$ for the same runs as in Fig.\ \ref{f1a} demonstrating the
proportionality $\Gamma \propto E$.} 
\label{f3}
\end{figure}

Also the time evolution of the energy ratio $\Gamma$ exhibits a uniform
behavior which 
is demonstrated in Fig.\ \ref{f2}. The slight shift of the 
uppermost curve corresponding to
the smallest value of $H$ (run 4), is due to the smaller drop of $\Gamma$ during the
very first phase of turbulence generation $t<0.5$ not included in the figure.
Moreover, we find that $\Gamma(t)$ is proportional to $E(t)$,
$\Gamma \simeq c E/H$, $c=0.1-0.15$,  as seen
in Fig.\ \ref{f3}, where $\Gamma/(E/H)$ is plotted.
Inserting this
result in Eq.\ (\ref{4}) we obtain the differential equation  for $E$,
which in the asymptotic limit $\Gamma \ll 1$ becomes
\begin{equation}
-\frac{dE}{dt} \simeq 0.5\frac{ E^3}{H^{3/2}}
\label{5}
\end{equation}
with the similarity solution $E\sim t^{-0.5}$. For finite 
$\Gamma$ the theory predicts a somewhat steeper decay flattening asymptotically
to $t^{-0.5}$ as $\Gamma$ becomes small, which is exactly the behavior of $E(t)$ observed 
in the simulations. (Note, that if $E(t)$ is plotted on the traditional 
log-log scale, which overrates the transition period $t\sim 1$, a steeper
decay would be suggested.) The relation $\Gamma \propto E$ now gives also the
similarity law for the kinetic energy $E^V \sim t^{-1}$.

This theory does not apply to the special case $H=0$. Here we find indeed a different
decay law, $E\sim t^{-1}$ from run 3, which is consistent with previous simulations
at lower Reynolds numbers \cite{8} and with the prediction in Ref.\ \cite{4}. The
transition to the slower decay for finite $H$ occurs at relatively small values,
0.1--0.2 of the maximum possible value. 

We have also studied the effect of an initial 
velocity and magnetic field alignment $\rho_0=K/E$. 
For small $\rho_0 < 0.1$ the alignment, after increasing initially, 
tends to saturate at some small value, which is due to the fact that $K$ is less well conserved
than $H$. For higher $\rho_0>0.3$ (runs 9 and 10 in Table I) 
the alignment becomes very strong, 
which as expected slows down the
energy decay drastically.

In conclusion we have presented a new phenomenology of the energy decay in 
3D incompressible MHD turbulence, which agrees very well with direct numerical simulations
at relatively high Reynolds numbers. We consider in particular the case of
finite magnetic helicity $H$, which is typical for naturally occuring magnetic
turbulence. The energy decay is governed by the conservation of $H$ and the
time evolution of the energy ratio $\Gamma=E^V/E^M$. We find that the relation
$(E^{5/2}/\epsilon H)\Gamma^{1/2}/(1+\Gamma)^{3/2} \simeq const$ is satisfied
for most $H$-values and is independent of the magnitude of the
dissipation coefficients and the order of the diffusion operator,
provided the Reynolds number is sufficiently high such that $H$ is well conserved.
The kinetic energy is found to decrease more rapidly than the magnetic one,
in contrast to the behavior in 2D, in particular we find $\Gamma \propto E$.
This proportionality leads to a simple energy decay law, $-dE/dt \sim E^3$,
or $E\sim t^{-0.5}$. We also
obtain the similarity law for the kinetic energy $E^V \sim t^{-1}$.
For the special case $H=0$ the energy decays more rapidly, $E\sim t^{-1}$,
which agrees with previous simulations at lower Reynolds numbers.
The transition to the finite-$H$ behavior occurs at relatively small
values of $H$.

Results concerning the spatial scaling properties of 3D 
MHD turbulence will be published in a subsequent
paper.

The authors would like to thank 
Andreas Zeiler for  providing the basic version of the code, Antonio Celani
for developing some of the diagnostics, 
and Reinhard Tisma for
optimizing the code for the CRAY T3E computer.

\narrowtext
\narrowtext
\narrowtext
\begin{table}
\caption{Summary of the simulation runs. The value of $H=0.28$ corresponds to the maximum
value for the given spectrum (\ref{2a}), $H\leq E/k_0$.}
\begin{center}
\begin{tabular}{ccccccc}
run No  &  $N$  &   $\nu$  &  $\eta_{\nu}$  &  $H$  &  $\rho_0$  &  $t_{\rm max}$ \\ \hline
   1  &     256   &     1   &     $10^{-3}$   &   0.19     &   0.04     &    18.5   \\
   2  &     512   &     1   &   $3\times 10^{-4}$ & 0.19   &   0.04     &    10  \\
   3  &     256   &     2   &     $10^{-6}$   &      0    &    0.05    &    20   \\
   4  &     256   &     2   &     $10^{-6}$   &     0.11  &    0.05    &    10   \\
   5  &     256   &     2   &     $10^{-6}$   &     0.19   &   0.04    &    20   \\
   6  &     512   &     2   &   $3\times 10^{-8}$ & 0.19   &   0.04     &    10   \\
   7  &     256   &     2   &     $10^{-6}$   &     0.25  &    0.04    &    10   \\  
   8  &     256   &     2   &     $10^{-6}$   &     0.28     & 0.03       &  10     \\
   9  &     256   &     2   &     $10^{-6}$   &     0.19    &  0.38      &    10   \\
   10 &     256   &     2   &     $10^{-6}$   &     0.19    &  0.71     &    10   \\
\end{tabular}
\end{center}
\label{t1}
\end{table}

\end{multicols}
\end{document}